\documentclass[a4paper,11pt,reqno]{amsart}
\usepackage{amsmath,amssymb,amsthm} 
\usepackage{graphics} \usepackage{enumerate}
\usepackage{enumitem}
\usepackage{dsfont}
\usepackage[colorlinks,citecolor=blue]{hyperref}
\usepackage[capitalize]{cleveref}
\usepackage{epsfig}
\usepackage{color}
\usepackage[english]{babel}

\numberwithin{equation}{section}

\newtheorem{theorem}{Theorem}[section]
\newtheorem{lemma}[theorem]{Lemma}
\newtheorem{proposition}[theorem]{Proposition}

\theoremstyle{definition}

\newtheorem{remark}[theorem]{Remark}

\newtheorem{example}[theorem]{Example}

\title{On distances among Slater Determinant States and Determinantal Point Processes
}
\author[C. Boccato]{Chiara Boccato}
\address{C.B.: Dipartimento di Matematica, Università degli Studi di Pisa, 56125 Pisa, Italy  }
\email{chiara.boccato@unipi.it}

\author[F. Pieroni]{Francesca Pieroni}
\address{F.P.: Dipartimento di Matematica, Guido Castelnuovo, Sapienza Universit\`a di Roma, 00185 Roma, Italy}
\email{francesca.pieroni@uniroma1.it}

\author[D. Trevisan]{Dario Trevisan}
\address{D.T.: Dipartimento di Matematica, Università degli Studi di Pisa, 56125 Pisa, Italy  }
\email{dario.trevisan@unipi.it}
\date{\today}

\begin{document}

\maketitle

\begin{abstract}
Determinantal processes provide mathematical modeling of repulsion among points. In quantum mechanics, Slater determinant states generate such processes, reflecting fermionic behavior. This note exploits the connections between the former and the latter structures by establishing quantitative bounds in terms of trace/total variation and Wasserstein distances.
\end{abstract}

\section{Introduction}

In quantum mechanics, Slater determinants provide an elegant representation of many-particle wavefunctions, encapsulating the Pauli exclusion principle that governs fermionic behavior \cite{piela2013ideas,BPS}. In classical probability, determinantal point processes have emerged as a fundamental stochastic model with applications spanning from statistical physics and random matrix theory \cite{borodin2009determinantal}, to machine learning  \cite{kulesza2012determinantal}. These processes are characterized by an intrinsic repulsion between points, making them both a valuable tool for modeling diverse phenomena wherenegative correlations play a key role, and a natural classical analogue of fermionic behavior. 

The link between these two perspectives has been well understood since the seminal work of Macchi \cite{macchi1975}, who observed that the squared modulus of a Slater determinant induces a determinantal point process whose kernel is the one-particle density matrix. This insight was later extended to more general quasi-free states—including thermal and ground states of quadratic Hamiltonians \cite{Soshnikov2000, Lytvynov2002, Gottlieb2005}, and systematically developed in \cite{Lyons2003,ShiraiTakahashi2003,Virag2006,hough2009zeros}. Thus, determinantal point processes naturally encode the spatial statistics of (effectively) free fermions, and provide convenient classical models for use in simulations. Their mathematical properties are by now well understood, yet quantitative relations between distances of quantum fermionic states and distances between the induced point process laws remain largely unexplored.


The goal of this note is to explore precisely an interplay between determinantal structures, quantum information theory, and optimal transport. Specifically, we examine how estimating distances between quantum Slater determinant states (\cref{prop:slater-W1}) give rise to upper bounds for the classical laws of determinantal point processes (\cref{thm:determinant-distance}), rectifying some bounds existing in the literature \cite{decreusefond2021optimal}. This is obtained by establishing rigorous connections between distances such as the trace and quantum Wasserstein distance of order $1$, and their classical counterparts,  providing quantitative control of the map
$$ \text{quantum fermionic state} \quad \mapsto \quad \text{classical determinantal point process law}.$$
Although we do not address specific applications in this note, we believe that our results could be naturally relevant for approximation schemes, stability questions, and classical simulation of fermionic systems.

The paper is organized as follows. In \cref{sec:general}, we provide an overview of the mathematical framework for distance measures between quantum states and their classical counterparts. In Section \ref{sec:w1-slater} we address the problem of bounding from above and below the quantum Wasserstein distance between Slater determinant states, and in Section \ref{sec:determinantal} we provide the application to bounds between laws of determinantal point processes.

\section{Generalities}\label{sec:general}

\subsection{Classical optimal transport} We refer to standard texts on optimal transport \cite{ambrosio2008gradient, villani2008optimal} for more details on the following basic facts. Given two Borel probability measures $\mu$, $\nu$ on a Polish space $E$, a lower-semicontinuous cost function $c: E\times E \to [0, \infty)$, the Kantorovich optimal transport problem is the minimization problem
\[
W_c(\mu, \nu):= \inf_{\pi \in \mathcal{C}(\mu, \nu)} \int_{E \times E} c(x,y) d \pi(x,y),
\]
where $\mathcal{C}(\mu, \nu)$ denotes the set of couplings between $\mu$ and $\nu$, i.e., probabilities on the product space $E\times E$ such that the first and second marginals are respectively $\mu$ and $\nu$. When $c(x,y)=d(x,y)$ is a distance function on $E$ and $\mu$, $\nu$ have finite first moments (with respect to the distance $d$), one can show that the $\inf$ in the transport problem is actually attained and the cost defines a distance on the set of such probabilities, also known as Wasserstein distance of order $1$. One can also establish the Kantorovich duality formula
\[W_c(\mu, \nu) = \sup \left\{\int_E f d \mu - \int_E fd \nu \, : \, |f(x) - f(y) | \le d(x,y) \quad \forall x,y \in E\right\}.
\]
In terms of notations, it will  be also convenient to write $W_c(X,Y)$ instead of $W_c(P_X, P_Y)$, whenever $X$ and $Y$ are random variables  with laws $P_X$ and $P_Y$ (and finite first moment).

\subsection*{Quantum systems}
A quantum system is represented by a complex Hilbert space $\mathcal{H}$. We use Dirac notation $\left|\psi\right \rangle \in \mathcal{H}$, $\left \langle\psi\right|\in \mathcal{H}^*$. In all our results, one can actually restrict to finite dimensional spaces. Quantum states are represented by state operators $\rho \in \mathcal{S}(\mathcal{H})$, i.e., self-adjoint nonnegative trace-class operators with ${\rm tr}[\rho]=1$. In finite dimension, quantum observables $\mathcal{O}(\mathcal{H})$ are represented by (bounded) self-adjoint operators $H$. As a generalization of the notion of quantum observable, we recall that a (sharp) measurement from the system $\mathcal{H}$ to a measurable space $E$, endowed with a $\sigma$-algebra $\mathcal{E}$ can be  defined by a Projection-Valued Measure (PVM) $\Pi$, so that $\Pi(A)$ is a normalized orthogonal projection on $\mathcal{H}$ for every $A \subseteq \mathcal{E}$ and $A \mapsto \Pi(A)$ is $\sigma$-additive on disjoint sets in the strong operator topology. Functional calculus on observables can be naturally extended to PVMs as follows: for any measurable bounded function $f: E \to \mathbb{R}$, one defines via ``composition'' the linear bounded operator $\Pi f$ on $\mathcal{H}$ via the bilinear form on $\mathcal{H}$,
\[
\left\langle\varphi\right|(\Pi f) \left|\psi\right \rangle =  \int_E f(x) \left \langle\varphi\right| \Pi(dx) \left|\psi\right \rangle.
\]
By duality, given a state operator $\rho \in \mathcal{S}(\mathcal{H})$, the push-forward of $\rho$ via $\Pi$ is the probability measure $\Pi_\sharp \rho$ on $E$ given by
\begin{equation*}
 \mathcal{E} \ni A \mapsto {\rm tr}[ \Pi(A) \rho].
\end{equation*}
In particular, one has the abstract change of variables
\[
\int_E f d \Pi_\sharp \rho = {\rm tr}[ (\Pi f) \rho  ].
\]
It is useful to associate to the PVM from $\mathcal{H}$ to $E$, when $\mathcal{H}$ is prepared on the state $\rho$, a classical random variable with values in $E$, that we denote with $X_\rho$, with law $\Pi_\sharp \rho$, so that the change of variables reads $\mathbb{E}\left[ f(X_\rho)\right] = {\rm tr}[ (\Pi f) \rho]$.

\begin{example}
Consider $H = L^2( E, \mathcal{E}, \mu)$, so that $\left|\psi\right \rangle = (\psi(x))_{x \in E}$ and the scalar product reads
\[
\left<\phi|\psi\right> = \int_{E} \overline{\phi(x)} \psi(x) d \mu(x).
\]
For $A \in \mathcal{E}$, let $\Pi(A)$ denote the multiplication operator by the indicator function of $A$, $\Pi(A)\left|\psi\right \rangle = \left|\chi_A \psi\right \rangle$. Then, one has that
\begin{equation}\label{eq:pif-phi-psi}  \left \langle\phi\right| (\Pi f) \left|\psi\right \rangle= \int_E f(x) \overline{\phi(x)} \psi(x) d \mu(x).\end{equation}
\end{example}

\subsection*{Trace distance}
On a quantum system described by a Hilbert space $\mathcal{H}$, the trace distance between two state operators $\rho$, $\sigma \in \mathcal{S}(\mathcal{H})$ is given by
\begin{equation*}
\left\|\rho - \sigma\right\|_{1} =\frac 1 2  {\rm tr} \left|\rho -\sigma\right| = \frac 1 2 {\rm tr} \left[\sqrt{ (\rho - \sigma)^2 }\right].
\end{equation*}
For pure states $\rho = \left|\psi\right \rangle\left \langle\psi\right|$, $\sigma = \left|\varphi\right \rangle\left \langle\varphi\right|$ associated to state vectors $\left|\psi\right \rangle$, $\left|\varphi\right \rangle\in \mathcal{H}$, the trace distance reduces to a function of the state overlap (or the fidelity):
\begin{equation}\label{eq:trace-pure}
\left\|\rho - \sigma\right\|_{1}  = \sqrt{ 1-  \left| \left<\psi | \varphi\right>\right|^2 }.
\end{equation}
One has the following duality formula for the trace distance
\begin{equation}\label{eq:trace-dual}
\left\| \rho -\sigma\right\|_1  = \sup_{0 \le H \le \mathds{1}} {\rm tr}[H (\rho - \sigma)],
\end{equation}
where $H \in \mathcal{O}\left(\mathcal{H}\right)$ and the inequalities are in the sense of quadratic forms, $\mathds{1}$ being the identity operator.

The trace distance is the quantum counterpart of the total variation distance between two probability measures $\mu$, $\nu$ on a measurable set $E$, which can be defined e.g.\ directly as a supremum over measurable functions
\[
\left\| \mu - \nu\right\|_{\operatorname{TV}} = \sup_{ f: E \to [0,1]} \int_E fd (\mu - \nu),
\]
or equivalently as the Wasserstein distance of order $1$ with respect to the trivial cost $d(x,y) = 1_{x\neq y}$.

As we are often dealing with random variables, it is convenient to denote the classical total variation distance between the laws $P_X$, $P_Y$ of two random variables $X$, $Y$ by specifying only the variables:
\[
\operatorname{TV}( X, Y) :=\left\|P_X - P_Y\right\|_{\operatorname{TV}} = \sup_{f: E \to [0,1]} \left\{\mathbb{E}\left[f(X)\right] - \mathbb{E}\left[f(Y)\right]\right\}.
\]
The trace distance contracts when performing any measurement on the quantum system $\mathcal{H}$ with outcome values in a (measurable) set $E$. Precisely, given any PVM $\Pi$ from $\mathcal{H}$ to $E$, the total variation distance between the push-forward  measures is bounded from above  by the total variation distance between  the quantum states:
\begin{equation}\label{eq:contraction-trace-measurement}\left\| \Pi_\sharp \rho - \Pi_\sharp \sigma\right\|_{\operatorname{TV}} \le\left\|\rho -\sigma\right\|_1.\end{equation}
This can be easily seen by the fact that for any measurable $f: E \to [0,1]$, the operator $H := \Pi f$ is self-adjoint with $0 \le H \le \mathds{1}$, and using the abstract change of variables. With the probabilistic notation $X_\rho$, $X_\sigma$, for the random variables associated to the PVM and the states $\rho$, $\sigma$, inequality \eqref{eq:contraction-trace-measurement} becomes
\begin{equation}\label{eq:contraction-trace-probabilities}
 \operatorname{TV}(X_\rho, X_\sigma) \le\left\|\rho - \sigma\right\|_1.
\end{equation}

\subsection*{Quantum Wasserstein distance of order 1}
The quantum Wasserstein distance of order $1$ on a product system $\mathcal{H} = \bigotimes_{i=1}^n \mathcal{H}_i$ was first introduced in \cite{de2021quantum}. Given state operators $\rho$, $\sigma \in \mathcal{S}(\mathcal{H})$, it is defined as follows:
\begin{equation*}\begin{split}
 &\left\|\rho - \sigma\right\|_{W_1} := \\
 & \min \left\{\sum_{i=1}^n c_i \, : \, \rho - \sigma = \sum_{i=1}^n c_i \left( \rho^{(i)} - \sigma^{(i)}\right) , \rho^{(i)}, \sigma^{(i)} \in \mathcal{S}(\mathcal{H}), \, {\rm tr}_i \rho^{(i) } = {\rm tr}_i \sigma^{(i)} , c_i\geq0\,\forall i \right\}\end{split}
\end{equation*}
where ${\rm tr}_i$ denotes the partial trace over the $i$-th sub-system $\mathcal{H}_i$. One can alternatively set $X_i := c_i (\rho^{(i)} - \sigma^{(i)})$ and minimize instead \begin{equation*}
  \sum_{i=1}^n\left\|X^{(i)}\right\|_{1}
\end{equation*}
among all the representations $\rho- \sigma = \sum_{i=1}^n X^{(i)}$ with $X^{(i)}$ self-adjoint and ${\rm tr}_i X^{(i)} = 0$ for every $i=1,\ldots, n$.
A dual formulation can be given as a supremum over quantum observables $H$,
\begin{equation}\label{eq:dual-QW1}
\left\|\rho - \sigma\right\|_{W_1}  = \sup_{\left\|H\right\|_{\operatorname{Lip}}\le 1} {\rm tr}[H (\rho - \sigma)],
\end{equation}
where the quantum Lipschitz constant of $H \in \mathcal{O}\left( \mathcal{H}^{\otimes n} \right)$ is defined as
\begin{equation}\label{eq:Q-Lip}
\left\|H\right\|_{\operatorname{Lip}} := 2 \max_{i=1,\ldots, n} \min_{H^{(i)}} \left\|H - \mathds{1}^{(i)}\otimes H^{(i)} \right\|_\infty,
\end{equation}
where $H^{(i)} \in \mathcal{O}\left( \bigotimes_{j\neq i} \mathcal{H}_j\right)$ denotes an observable on the product obtained by removing the $i$-th system and $\mathds{1}^{(i)}$ denotes the identity operator on $\mathcal{H}_i$. Choosing $H^{(i)} = 0$, it follows that $\left\|H\right\|_{\operatorname{Lip}}\le 2\left\|H\right\|_\infty$  hence by duality one obtains that
\begin{equation}\label{eq:w-1-lower-tr}
\left\|\rho - \sigma\right\|_{W_1} \ge\left\|\rho - \sigma\right\|_{1}.
\end{equation}
 One has also the inequality
\begin{equation}
 \label{eq:upper-bound-w1-tr}
 \frac 1 n\left\|\rho - \sigma\right\|_{W_1}  \le\left\|\rho - \sigma\right\|_{1}.
\end{equation}
The quantum Wasserstein distance of order $1$ is the quantum analogue of the optimal transport cost between classical probabilities on a product set $E = \prod_{i=1}^nE_i$, when computed with respect to the Hamming distance, i.e., for $x, y \in E$,
\begin{equation*}
 d_{H}(x,y) = \sum_{i=1}^n 1_{\left\{x_i \neq y_i\right\}}.
\end{equation*}
This can be conveniently seen by comparing the dual formulation  \eqref{eq:dual-QW1} with the classical Kantorovich dual formula, for $\mu$, $\nu$ probabilities on $E^n$,
\begin{equation*}
 W_1\left(\mu, \nu\right) := \sup_{\operatorname{Lip}(f) \le 1} \int_{E^n} f d (\mu - \nu),
\end{equation*}
where $\operatorname{Lip}(f)$ denotes the Lipschitz constant with respect to the Hamming distance. It is easy to notice that $\operatorname{Lip}(f)\le 1$ if and only if  $|f(x)-f(x')| \le 1$ whenever $x, x'\in E$ differ only for a single coordinate, and similarly that the following classical analogue of \eqref{eq:Q-Lip} holds:
\begin{equation*}
\left\|f\right\|_{\operatorname{Lip}} := 2 \max_{i=1,\ldots, n} \min_{f^{(i)}} \left\|f - f^{(i)}\right\|_\infty,
\end{equation*}
where each function $f^{(i)}$ does not depend on the $i$-th coordinate, i.e., $f^{(i)}: \prod_{j\neq i}E_j \to \mathbb{R}$

An analogue of the contraction property \eqref{eq:contraction-trace-measurement} holds for the quantum Wasserstein distance of order $1$, when performing $n$ measurements separately on the subsystems. Precisely, given $(\Pi_i)_{i=1}^n$, where $\Pi_i$ is a PVM on $\mathcal{H}_i$, with outcome values in a (measurable) set $(E_i, \mathcal{E}_i)$, one can define a joint PVM $\Pi = \bigotimes_{i=1}^n \Pi_i$ on $\mathcal{H}^{\otimes n}$, taking values in $\prod_{i=1}^n E_i$: first, on rectangle sets $A = \prod_{i=1}^n A_i$, we set $\Pi(A) = \bigotimes_{i=1}^n \Pi_i(A_i)$ and we then extend it on  the product $\sigma$-algebra $\bigotimes_{i=1}^n \mathcal{E}_i$. Then,  it holds
\begin{equation}\label{eq:contraction-W-measurement} W_1(\Pi_\sharp \rho, \Pi_\sharp \sigma) \le\left\|\rho-\sigma\right\|_{W_1}.\end{equation}
Indeed, with the same notation of the previous section, one has that for any measurable $f$ on $E$ with $\operatorname{Lip}(E) \le 1$, the operator $H = f \circ \Pi$ has quantum Lipschitz constant $\left\|H\right\|_{\operatorname{Lip}} \le 1$, since one can use $H^{(i)} := f^{(i)} \circ (\bigotimes_{j\neq i} \Pi_j)$ in \eqref{eq:Q-Lip}.

\begin{remark}
Using the notation for random variables, we notice that to each PVM $\Pi_i$ we can associate a random variable $X_{\rho,i}$ with values in $E_i$ (assuming that the system $\mathcal{H}$ is in state $\rho$, so that the joint random variable $X_\rho$ associate to the joint PVM $\Pi$ takes values in $E$ and its marginal variables are (equal in law to) $(X_{\rho,i})_{i=1}^n$. Hence, we write $X_\rho = (X_{\rho,i})_{i=1}^n$. Similarly, we can write $X_{\sigma} = (X_{\sigma, i})_{i=1}^n$, and \eqref{eq:contraction-W-measurement} reads
\begin{equation}\label{eq:contraction-W-random-variables}
 W_1( (X_{\rho, i})_{i=1}^n, (X_{\sigma, i})_{i=1}^n ) \le\left\|\rho - \sigma\right\|_{W_1},
\end{equation}
where we recall the small abuse of notation $W_1(X,Y) := W_1(P_X, P_Y)$ for the (Hamming-) Wasserstein distance between the laws of two random variables $X$, $Y$.
\end{remark}

\begin{remark}
It is actually known that
\[
 \sup_{\Pi} \operatorname{TV}(X_\rho, X_\sigma) =\left\|\rho - \sigma\right\|_{1},
 \]
where the supremum runs over all PVM on a system $\mathcal{H}$. However, the inequality
\[
\sup_{(\Pi_i)_{i=1}^n}   W_1( (X_{\rho, i})_{i=1}^n, (X_{\sigma, i})_{i=1}^n ) \le\left\|\rho - \sigma\right\|_{W_1},
\]
 may be strict in general. Consider for example on a two-qubits system $\mathcal{H} = (\mathbb{C}^2)^{\otimes 2}$, the maximally mixed state $\rho = \frac 1  4 \mathds{1} $, and $\sigma = \left|\psi\right \rangle\left \langle\psi\right|$ the uniform superposition pure state $\left|\psi\right \rangle= \frac 1 2\sum_{i,j=0}^1 \left|i\right \rangle\otimes\left|j\right \rangle$. Then for every $(\Pi_i)_{i=1}^2$, one has $W_1( (X_{\rho, i})_{i=1}^2, (X_{\sigma, i})_{i=1}^2 ) = 0$ since  the variables have the same law (in particular, all the marginals are independent), while $\left\|\rho - \sigma\right\|_{W_1} \ge \left\|\rho - \sigma\right\|_{1} >0$.
\end{remark}

\section{Wasserstein distance between Slater determinant states}\label{sec:w1-slater}

 On a quantum system $\mathcal{H}$, given orthonormal vectors $ \left\{\left|\psi_i\right \rangle \right\}_{i=1}^n \in \mathcal{H}$, one defines the Slater determinant state vector
\begin{equation*}
\left|\Psi\right \rangle = \frac 1 {\sqrt{n!}}\sum_{\tau \in \mathfrak{S}_n} (-1)^{\tau} \bigotimes_{i=1}^n \left|\psi_{\tau(i)}\right \rangle \in \mathcal{S}(\mathcal{H}^{\otimes n}),
\end{equation*}
where $\mathfrak{S}_n$ denotes the group of permutations over $n$ elements and $(-1)^\tau$ denotes the sign of the permutation $\tau$. It is well-known that $\left|\Psi\right \rangle$ defines a state vector, i.e., $\left\|\Psi\right\| = 1$, and more generally the overlap between the Slater determinant state vector $\left|\Psi\right \rangle$, associated to $\left\{\left|\psi_i\right \rangle \right\}_{i=1}^n$, and $\left|\Phi\right \rangle$,  associated to orthonormal vectors $\left\{\left| \phi_i\right \rangle \right\}_{i=1}^n$, is given by
\begin{equation*}
|\left<\Psi, \Phi\right>|^2 = \left| \det\left( \left<\psi_i | \phi_j\right>_{i,j=1\ldots, n}\right)\right|^2.
\end{equation*}
Therefore, their trace distance is given by \eqref{eq:trace-pure}, which reads
\begin{equation}\label{eq:slater-trace-distance}
\left\|\left|\Psi\right \rangle\left \langle\Psi\right| - \left|\Phi\right \rangle \left \langle\Phi\right\|\right|_{1} = \sqrt{ 1- \left| \det\left(\left<\psi_i | \phi_j\right>_{i,j=1\ldots, n}\right)\right|^2}.
\end{equation}
Let us notice that
\begin{equation*}
 \begin{split} \left| \det\left( \left<\psi_i | \phi_j\right>_{i,j=1\ldots, n}\right)\right|^2 & = \det\left(\left<\psi_i | \phi_j\right>_{i,j=1\ldots, n} \right) \det\left( \left<\phi_j | \psi_i\right>_{i,j=1\ldots, n}\right)\\
 & =\det\left(\left(\sum_{\ell=1}^n\left<\psi_i | \phi_\ell\right> \left<\phi_\ell | \psi_j\right>\right)_{i,j=1\ldots, n}\right)\\
 & = \det{\left( \left<\psi_i | K_\Phi \psi_j\right>\right)_{i,j=1,\ldots,n}},
 \end{split}
 \end{equation*}
where we define
\begin{equation}\label{eq:k-phi} K_\Phi = \sum_{\ell=1}^n \left|\phi_\ell\right \rangle \left \langle\phi_\ell\right|,\end{equation}
i.e. the orthogonal projection on the span of $\left\{\phi_\ell\right\}_{\ell=1,\ldots,n}$. Let us also introduce the stabilizer
\[
\mathfrak{U}_\Phi: = \left\{U: \mathcal{H} \to \mathcal{H}\, : \, \text{$U$ unitary such that  $U K_\Phi U^\dagger = K_\Phi$,} \right\},
\]
i.e., $U\left|\phi_i\right \rangle\in \operatorname{span}\left\{\left|\phi_j\right \rangle\right\}_{j=1, \ldots, n}$ for every $i=1,\ldots, n$. Then, for every $U \in \mathfrak{U}_\Phi$, setting $\psi_i = U \phi_i$, we have that $\Psi := U^{\otimes n} \Phi = \Phi$.  Indeed,
\[
\det{\left(\left<\psi_i | K_\Phi \psi_j\right>\right)_{i,j=1,\ldots,n}} = \det{\left(\left<\phi_i | \phi_j\right>\right)_{i,j=1,\ldots,n}} = \det\left( \mathds{1}_{\mathbb{C}^n}\right) = 1,
\]
and therefore
\[
\left\|\left|\Psi\right \rangle\left \langle\Psi\right| - \left|\Phi\right \rangle \left \langle\Phi\right|\right\|_{1} = 0.
\]
Let us also notice that $K_\Phi$ is, up to a factor $1/n$, the reduced density operator of $\left|\Phi\right \rangle\left \langle\Phi\right|$ over any single system $\mathcal{H}$ (e.g., the first one):
\[
\frac 1 n K_\Phi =  {\rm tr}_{2,\ldots, n}[\left|\Phi\right \rangle\left \langle\Phi\right|].
\]
In fact, in quantum many-body systems, it is customary to define the $k$-particle  reduced density matrices
\begin{equation} \label{eq:rdm}\Gamma^{(k)}_\Phi:= {n \choose k}{\rm tr}_{(k+1),\ldots, n}[\left|\Phi\right \rangle\left \langle\Phi\right|],
\end{equation}
so that $K_\Phi = \Gamma^{(1)}_{\Phi}$.

In the next result, we provide upper and lower bounds for the quantum Wasserstein distance of order $1$ between Slater determinant states.
\begin{theorem}\label{prop:slater-W1}
Given Slater determinant state vectors $\left|\Psi\right \rangle$, $\left|\Phi\right \rangle$ associated respectively to orthonormal vectors $\left\{\left|\psi_i\right \rangle\right\}_{i=1}^n$, $\left\{\left|\phi_i\right \rangle\right\}_{i=1}^n  \subseteq \mathcal{H}$, it holds
\begin{equation}\label{eq:wass-slater}
\left\|\left|\Psi\right \rangle\left \langle\Psi\right| - \left|\Phi\right \rangle\left \langle\Phi\right| \right\|_{W_1} \le n \sqrt{ 1 - \max_{V \in \mathfrak{U}_{\Psi}, U \in \mathfrak{U}_\Phi}\left| \frac 1 n \sum_{i=1}^n \left<V \psi_{i} |  U \phi_{i} \right>\right|^2 }.
\end{equation}
Moreover, the sequence
\begin{equation} \label{eq:wass-rdm} k \mapsto \frac 1 k \left\|{n \choose k}^{-1} \Gamma^{(k)}_\Phi -{n \choose k}^{-1}  \Gamma^{(k)}_\Psi\right\|_{W_1}\end{equation}
is non-decreasing for $k=1, \ldots, n$.
\end{theorem}

We currently are not able to characterize the cases when equality holds in \eqref{eq:wass-slater}, see also Section  \ref{sec:conclusion} for a list of open questions.

\begin{proof}
To show  \eqref{eq:wass-slater},  one can assume without loss of generality that the optimal $V \in \mathfrak{U}_\Psi$, $U \in \mathfrak{U}_\Phi$ in \eqref{eq:wass-slater} are the identity map, up to replacing the given families with  $\left\{V\left|\psi_i\right \rangle\right\}_{i=1}^n$ and $\left\{U\left|\phi_i\right \rangle\right\}_{i=1}^n$ which does not change the state vectors $\Phi$, $\Psi$.

Let then  $U$ denote a unitary map on $\mathcal{H}$ such that $U\left|\psi_i\right \rangle = \left|\phi_i\right \rangle$ for every $i=1, \ldots, n$. For every $k=0, \ldots, n$ consider the state vector
\begin{equation*} \left|\Psi_k\right \rangle:= (U^{\otimes k} \otimes \mathds{1}_{\mathcal{H}}^{\otimes (n-k)}) \left|\Psi\right \rangle= \frac 1 {\sqrt{n!}} \sum_{\tau \in \mathfrak{S}_n} (-1)^\tau \bigotimes_{i=1}^k \left|\phi_{\tau(i)}\right \rangle  \otimes \bigotimes_{i=k+1}^n \left|\psi_{\tau(i)}\right \rangle
\end{equation*}
and the associated state operator $\rho_k = \left|\Psi_k\right \rangle \left \langle\Psi_k\right|$.

Since $\left|\Psi_0\right \rangle= \left|\Psi\right \rangle$ and $\left|\Psi_n\right \rangle = \left|\Phi\right \rangle$, we have
\[
\left|\Psi\right \rangle\left \langle\Psi\right|- \left|\Phi\right \rangle\left \langle\Phi\right|= \sum_{k=1}^n \rho_{k-1}-\rho_k
\]
and since
\[
\left|\Psi_k\right \rangle = \mathds{1}_{\mathcal{H}}^{\otimes (k-1)} \otimes U \otimes \mathds{1}_{\mathcal{H}}^{\otimes (n-k)} \left|\Psi_{k-1}\right \rangle,
\]
 i.e., $\rho_k$ can be obtained from $\rho_{k-1}$ by acting only on the $k$-th subsystem, we have
\begin{equation*}
 {\rm tr}_k \rho_k = {\rm tr}_k \rho_{k-1}.
\end{equation*}
Thus, we find
\begin{equation*}
\left\|\left|\Psi\right \rangle\left \langle\Psi\right|- \left|\Phi\right \rangle\left \langle\Phi\right| \right\|_{W_1} \le \sum_{k=1}^n \left\| \rho_k - \rho_{k-1}\right\|_{1} = \sum_{k=1}^n \sqrt{1- \left|\left<\Psi_k|\Psi_{k-1}\right>\right|^2 },
\end{equation*}
where the equality follows because the states operators $\rho_{k-1}$, $\rho_k$ are pure. Moreover, such overlap does not depend on $k$, and it equals
\begin{equation*}\begin{split}
\left|\left<\Psi_k|\Psi_{k-1}\right>\right|^2 & =
\left|\frac{1}{n!} \sum_{\tau, \tau'} (-1)^\tau(-1)^{\tau'}\prod_{i=1}^{k-1} \left<\phi_{\tau(i)}| \phi_{\tau'(i)}\right> \cdot  \left<\phi_{\tau(k)} |  \psi_{\tau(k)}\right>\cdot  \prod_{i=k+1}^n\left< \psi_{\tau(i)}| \psi_{\tau'(i)}\right>\right|^2 \\
& =\left|\frac{1}{n!} \sum_{\tau} \left<\phi_{\tau(k)} | \psi_{\tau(k)}\right> \right|^2\quad \text{(for the cases $\tau\neq \tau'$ yield no contribution)}\\
& = \left| \frac 1 n \sum_{i=1}^n \left<\phi_{i} | \psi_i\right>\right|^2
\end{split}
\end{equation*}
since there are $(n-1)!$ permutations $\tau$ with fixed $\tau(k) = i$, for each $i=1, \ldots, n$.

To prove \eqref{eq:wass-rdm}, given $H \in \mathcal{O}(\mathcal{H}^{\otimes k})$ with $\left\| H\right\|_{\operatorname{Lip}} \le 1$, consider the observable
\[
\tilde H = \sum_{j=1}^{k+1} \mathds{1}^{(j)} \otimes H _j \in \mathcal{O}(\mathcal{H}^{\otimes (k+1)}),
\]
 where we denote with $\mathds{1}_{j}$ the identity on the $j$-th site, and $H_j \in \mathcal{O}(\mathcal{H}^{\otimes k})$ denotes a 'copy' of $H$ that sits on the $k$ subsystems after we remove the $j$-th copy. We claim that the quantum Lipschitz constant of $\tilde H$ is (less than) $k$, so that
 \begin{equation*}
 \begin{split}
  (k+1) {\rm tr}[ H & \left({\rm tr}_{(k+1),\ldots, n}[\left|\Psi\right\rangle\left \langle\Psi\right|] - {\rm tr}_{(k+1),\ldots, n}[\left|\Phi\right \rangle\left \langle\Phi\right|]\right) ]\\
   & = \sum_{j=1}^{k+1} {\rm tr}[ H_j {\rm tr}_{j, k+2,\ldots, n}[\left|\Psi\right \rangle\left \langle\Psi\right|- \left|\Phi\right \rangle\left \langle\Phi\right|] ]  \\
 & = {\rm tr}[ \tilde H {\rm tr}_{k+2, \ldots, n}[\left|\Psi\right \rangle\left \langle\Psi\right| - \left|\Phi\right \rangle\left \langle\Phi\right| ]\\
& \le k \left\|{\rm tr}_{k+2, \ldots, n}[\left|\Psi\right \rangle\left \langle\Psi\right| ] -{\rm tr}_{k+2, \ldots, n}[\left|\Phi\right \rangle\left \langle\Phi\right| ]  \right\|_{W_1}.
 \end{split}
 \end{equation*}
Taking supremum over $H$ and recalling  \eqref{eq:rdm}, we obtain
\[
\frac 1 k \left\| {n \choose k}^{-1} \Gamma^{(k)}_\Phi -{n \choose k}^{-1}  \Gamma^{(k)}_\Psi\right\|_{W_1} \le  \frac 1 {k+1} \left\|{n \choose k+1}^{-1} \Gamma^{(k+1)}_\Phi -{n \choose k+1}^{-1}  \Gamma^{(k+1)}_\Psi\right\|_{W_1}.
\]
To argue that $\left\|\tilde H\right\|_{\operatorname{Lip}} \le k$, consider any $i \in \left\{1, \ldots, k+1\right\}$, take $H^{(i)}$ such that
\[
2 \left\|H - H^{(i)}\otimes \mathds{1}^{(i)}\right\|_\infty \le 1,
\]
and set
\[
\tilde H^{(i)} = H_i + \sum_{j\neq i} H_j^{(i)}\otimes \mathds{1}^{(j)},
\]
where we use the same notation as for $H$, i.e. $H_j^{(i)}$ is a copy of $H^{(i)}$ on the system obtained by removing both sites $i$ and $j$. Then,
\[
2\left\|\tilde H - \tilde H^{(i)}\otimes \mathds{1}^{(i)}\right\|_\infty \le 2\left\|\sum_{j\neq i}\left(H_j- H_j^{(i)} \right) \right\|_\infty \le k. \qedhere
\]
 \end{proof}

\begin{example}
In view of \eqref{eq:upper-bound-w1-tr}, it is reasonable to compare the upper bound between Slater determinant state operators provided by \eqref{eq:wass-slater} with $n$ times their trace distance. The following example shows that it can hold indeed, as $n \to \infty$,
\[
 \sqrt{ 1 - \max_{V \in \mathfrak{V}_\Psi, U \in \mathfrak{U}_\Phi}\left| \frac 1 n \sum_{i=1}^n \left<V \psi_{i} | U\phi_{i}\right> \right|^2}\ll \sqrt{ 1- \left| \det\left(\left<\psi_i | \phi_j\right>_{i,j=1\ldots, n}\right)\right|^2}
 \]
so that the Wasserstein distance (divided by $n$) is much smaller than the trace distance: consider orthonormal vectors $\left\{\left|\psi_i\right \rangle\right\}_{i=1}^{2n}$, and set
\[
\left|\phi_i\right \rangle := (1-\varepsilon_i) \left|\psi_i\right \rangle + \sqrt{1- (1-\varepsilon_i)^2} \left|\psi_{n+i}\right \rangle, \quad \text{for $i=1, \ldots, n$,}
\]
where $(\varepsilon_i)_{i=1}^\infty \in [0,1]$ is such that $\sum_{i=1}^n \varepsilon_i < \infty$. Then,
\[
\det\left( \left<\psi_i| \phi_j\right>_{i,j=1,\ldots, n}\right) = \prod_{i=1}^n (1-\varepsilon_i) \stackrel{n\to \infty}{\to} \prod_{i=1}^\infty(1-\varepsilon_i) \in (0,1),
\]
while, choosing $U = V$ both equal to the identity, we find
\[
\frac 1 n \sum_{i=1}^n \left<\psi_{i} | \phi_{i}\right>= 1 - \frac 1 n \sum_{i=1}^n \varepsilon_i \stackrel{n\to \infty}{\to }1.
\]
\end{example}

\section{Application to determinantal point processes}\label{sec:determinantal}

Let $(E, \mathcal{E}, \mu)$ be a measure space, with $E$ Polish and $\mathcal{E} = \mathcal{B}(E)$ its Borel $\sigma$-algebra, and $\mu$ Radon. Given a measurable (complex valued) kernel $K = (K(x,y))_{x,y \in E}$ that induces an operator on $L^2(E, \mu)$ that is bounded self-adjoint, locally trace class and such that $0 \le K \le \mathds{1}$, one one can show \cite[section 4.5]{hough2009zeros} that there exists a determinantal point process $\mathcal{X}$, i.e., a random variable taking values in locally finite subsets of $E$ such that its joint intensities (also known as correlation functions) are, for every $m \ge 1$,
\[
\rho_m(x_1, \ldots, x_m) = \det\left( K(x_i, x_j)_{i=1, \ldots, m}\right).
\]
By definition, it means that for every $B_1, \ldots, B_k \in \mathcal{E}$ relatively compact and disjoint, it holds
\[
\mathbb{E}\left[ \prod_{i=1}^k { \sharp (\mathcal{X} \cap B_i) \choose m_i } m_i! \right] = \int_{B_1^{m_1}\times \ldots, \times B_k^{m_k}} \det\left( K(x_i, x_j)_{i=1, \ldots, m}\right) d \mu(x_1)\ldots, d \mu(x_m),
\]
where $m = \sum_{i} m_i$.
This entails in particular that, for every relatively compact $B \in \mathcal{E}$, it holds
\begin{equation}\label{eq:trace-determinant}\mathbb{E}\left[ \sharp (\mathcal{X} \cap B) \right] = \int_B \rho_1(x) d \mu(x) = {\rm tr}\left[ \chi_B K \chi_B \right].\end{equation}

The kernel $K$ (and the reference measure $\mu$, that we keep fixed in what follows) identifies the law of $\mathcal{X}$ (although the converse is not true). Therefore, one should be able to argue that if two kernels $K$, $K'$ are suitably close, then the associated determinantal point processes $\mathcal{X}$, $\mathcal{X}'$ are close in law.

Starting from the spectral decompositions
\[
K = \sum_{i=1}^\infty \lambda_i \left|\psi_i\right \rangle\left \langle\psi_i\right|, \quad K' = \sum_{i=1}^\infty \lambda_i' \left|\psi_i'\right \rangle\left \langle\psi_i'\right|,
\]
in \cite{decreusefond2021optimal} two results are given, providing quantitative bounds in two cases. In the first case, \cite[theorem 13]{decreusefond2021optimal}, assuming that all the eigenfunctions coincide, i.e., $\left|\psi_i\right \rangle=\left|\psi_i'\right \rangle$ for every $i$, a suitable Wasserstein distance between the laws is bounded in terms of the differences of the  eigenvalues $\lambda_i - \lambda_i'$. In the second case, assuming that all the eigenvalues coincide and that both $K$ and $K'$ are projection operators with same rank $n$, i.e., $\lambda_i=\lambda'_i=1$ for $i= 1, \ldots, n$, $\lambda_i=\lambda_i'=0$ for $i>n$, in \cite[Theorem 18]{decreusefond2021optimal}, the following inequality is claimed:
 \begin{equation}
 \label{eq:decreusefond-wrong} W_c\left( P_{\mathcal{X}}, P_{\mathcal{X}'} \right) \le \min_{\tau \in \mathfrak{S}_n} \sum_{i=1}^n W_2\left( |\psi_i|^2\mu, |\psi_{\tau(i)}'|^2\mu\right),
 \end{equation}
 where $W_2^2$ is the squared Wasserstein distance of order $2$ with respect to the Euclidean distance cost (here $E = \mathbb{R}^d$) and $W_c$ denotes the optimal transport cost defined with respect to the cost between subsets of $n$ points given by the minimum assignment problem with respect to the squared Euclidean distance cost:
\[
c(\left\{x_i\right\}_{i=1}^n, \left\{y_j\right\}_{j=1}^n ) = \min_{\sigma \in \mathfrak{S}_n} \sum_{i=1}^n |x_i- y_{\sigma(j)}|^2.
\]
 Unfortunately, such an inequality cannot hold, for the left hand side in \eqref{eq:decreusefond-wrong} is zero if and only if $\mathcal{X}$ equals $\mathcal{X}'$ in law, but in the right hand side the amplitudes $\left|\left<\psi_i|\psi_{j}\right>\right|^2$ for $i \neq j$ (and similarly for $\psi'$) do not appear. It is possible indeed to provide examples of pairs of determinantal point processes for which the right hand side is zero but the laws are different:  consider the case $n=2$, $E = [0,1]$ $\mu$ the Lebesgue measure, and starting from the first three Walsh functions $(w_0)_{i=0}^3$, set $\psi_1:=w_0$, $\psi_2:=w_1$, $\psi_1' := w_0$, $ \psi_2':=w_2$, so that one has identically $\left|\psi_i\right| = \left|\psi_i'\right| = 1$ and therefore the right hand side in \eqref{eq:decreusefond-wrong} vanishes, but $\mathcal{X}$ and $\mathcal{X}'$ have different laws, because an explicit computation gives 
\[
\operatorname{Cov}(\sharp (\mathcal{X} \cap (0,1/4)), \sharp ( \mathcal{X} \cap (1/4, 1/2) ) = -1/4,  \quad \operatorname{Cov}( \sharp (\mathcal{X}' \cap (0,1/4)), \sharp ( \mathcal{X}' \cap (1/4, 1/2)) ) = 0.
\]
Our first result in this section provides two alternative bounds, by measuring the distance between the laws of two determinantal processes with projection kernel operators in terms of their total variation or the Wasserstein distance, built with respect to the cost
\begin{equation*}
\# \left( A, B\right) := \sharp \left( A \Delta B\right)
\end{equation*}
for (locally) finite sets $A, B \subseteq E$. Such optimal transport cost between random variables taking values in finite sets is denoted with $W_{\operatorname{TV}}$ in \cite{decreusefond2021optimal} as it coincides with the total variation distance of the unnormalized empirical measures $\mu_A =\sum_{x \in A} \delta_x$, $\mu_B = \sum_{x \in B} \delta _x$.

\begin{proposition}\label{prop:determinant-distance}
Let $\mathcal{X}$, $\mathcal{X}'$ be determinantal processes on the same state space $E$, with common intensity measure $\mu$ and kernels $K$, $K'$ that are projection operators with same rank $n$, i.e.,
\begin{equation}\label{eq:kernel-kk-prime} K = \sum_{i=1}^n \left|\psi_i\right \rangle\left \langle\psi_i\right|, \quad K' = \sum_{i=1}^n \left|\psi_i'\right \rangle\left \langle\psi_i'\right|.
 \end{equation}
Then,
\begin{equation}\label{eq:tv-determinantal}
 \operatorname{TV}(\chi, \chi') \le \sqrt{ 1- \left| \det\left(\left<\psi_i | \psi_j'\right>_{i,j=1\ldots, n}\right)\right|^2},
\end{equation}
and
\begin{equation}\label{eq:w1-determinantal}
 W_{\#}\left( \chi, \chi'\right)\le  n \sqrt{ 1 - \max_{V \in \mathfrak{U}_\Psi, U \in \mathfrak{U}_{\Psi'}}\left| \frac 1 n \sum_{i=1}^n\left<V \psi_{i} | U\psi_{i}'\right> \right|^2},
\end{equation}

\end{proposition}

The key observation is that by a suitable joint measurement of the quantum system prepared $L^2(E^n, \mu^{\otimes n})$ on a Slater determinant state the resulting random variable yields a determinantal point process, up to forgetting the order of the measurements. 

\begin{lemma}\label{lem:slater-measure}
Let $\mathcal{H} = L^2(E,\mu)$ and consider a Slater determinant state operator $\rho =\left|\Psi\right \rangle \left \langle\Psi\right|\in \mathcal{S}(\mathcal{H}^{\otimes n})$ associated to  orthonormal vectors $\left\{\left|\psi_i\right \rangle\right\}_{i=1}^n\subseteq \mathcal{H}$, and the (joint) PVM on $\mathcal{H}^{\otimes n}$ given by the multiplication operation $\Pi(A)\left|\psi\right \rangle = \left|\chi_A \psi\right \rangle$ for $\psi \in L^2(E^n, \mu^{\otimes n})$, $A \in \mathcal{E}^{\otimes n}$.  Then, the random variable $\mathcal{X}_\rho := \left\{X_{\rho, i}\right\}_{i=1}^n$ is a determinantal point process with kernel
\[
K_\rho  (x,x') = \sum_{\ell=1}^n \overline{\psi_\ell}(x) \psi_\ell(x').
\]
\end{lemma}

\begin{remark}
Results closely related to the above are classical in the theory of determinantal point processes, see e.g.\ \cite[Lemma 4.5.1]{hough2009zeros}, where they are expressed in terms of the projection kernel governing the determinantal point process. We include a self-contained proof here because, in our setting, the statement is naturally phrased using projection-valued measures associated with the underlying one-particle operator and the Slater determinant structure of the quantum fermionic state. This formulation makes explicit the connection between the fermionic many-body description and the induced determinantal law, and it will be used later to relate quantum distances to classical Wasserstein bounds.
\end{remark}

\begin{proof}
We notice first that, for bounded measurable $f: E^n \to \mathbb{R}$ it holds, by definition of Slater determinant state and \eqref{eq:pif-phi-psi} applied on $E^n$, with measure $\mu^{\otimes n}$,
\begin{equation}\label{eq:exp-slater-f}
\begin{split}\mathbb{E}\left[ f( X_\rho)\right] & =\left<\Psi | (\Pi f) \Psi\right>\\
& = \frac 1 {n!}\sum_{\tau, \tau' \in \mathfrak{S}_n} (-1)^\tau (-1)^{\tau'} \bigotimes_{i=1}^n\left \langle\psi_{\tau(i)}\right| (\Pi f)  \bigotimes_{i=1}^n \left|\psi_{\tau'(j)}\right \rangle \\
& = \frac 1 {n!}\sum_{\tau, \tau' \in \mathfrak{S}_n} (-1)^\tau (-1)^{\tau'} \int_{E^n} f(x_1, \ldots, x_n) \prod_{i=1}^n \overline{\psi_{\tau(i)}}(x_i) \psi_{\tau'(i)}(x_i) d\mu(x_i).
\end{split}
\end{equation}
Let us first notice that the law of $X_\rho$ is exchangeable, i.e., for every permutation $\sigma \in \mathfrak{S}_n$, $(X_{\rho, \sigma(i)})_{i=1}^n$ has the same law as $X_\rho$. Indeed,
\begin{equation*}
\begin{split}& \mathbb{E}\left[f( (X_{\rho, \sigma(i)})_{i=1}^n )\right]  = \frac 1 {n!}\sum_{\tau, \tau' \in \mathfrak{S}_n} (-1)^\tau (-1)^{\tau'} \int_{E^n} f(x_{\sigma(1)}, \ldots, x_{\sigma(n)}) \prod_{i=1}^n \overline{\psi_{\tau(i)}}(x_i) \psi_{\tau'(i)}(x_i) d\mu(x_i)\\
 &\quad =\frac 1 {n!}\sum_{\tau, \tau' \in \mathfrak{S}_n} (-1)^\tau (-1)^{\tau'} \int_{E^n} f(x_{1}, \ldots, x_{n}) \prod_{i=1}^n \overline{\psi_{\tau \circ \sigma^{-1}(i)}}(x_i) \psi_{\tau'\circ \sigma^{-1}(i)}(x_i) d\mu(x_i) \\
&\quad  = \frac 1 {n!}\sum_{\tau, \tau' \in \mathfrak{S}_n} (-1)^\tau (-1)^{\tau'} \int_{E^n} f(x_{1}, \ldots, x_{n} ) \prod_{i=1}^n \overline{\psi_{\tau}(i)}(x_i) \psi_{\tau'(i)}(x_i) d\mu(x_i),
\end{split}
\end{equation*}
where in the last line we used the fact that $(-1)^{\tau \circ \sigma^{-1} } = (-1)^{\tau}(-1)^{\sigma}$, and similarly for $\tau'$, hence the sign of $\sigma$ cancels out.

Next, we argue that the process is simple, i.e., for any $i \neq j$, it holds $X_{\rho, i} \neq X_{\rho, j}$ a.s., hence in particular $\sharp \chi_\rho = n$. Indeed, choosing e.g.\ $f(x_1, \ldots, x_n) = 1_{x_1=x_1}$ -- without loss of generality, since $X_{\rho}$ is exchangeable -- we find
\begin{equation*}\begin{split}
P( X_{\rho, 1} = X_{\rho,2}) &= \frac 1 {n!}\sum_{\tau, \tau' \in \mathfrak{S}_n} (-1)^\tau (-1)^{\tau'} \int_{E^n} 1_{x_1=x_2} \prod_{i=1}^n \overline{\psi_{\tau(i)}}(x_i) \psi_{\tau'(i)}(x_i) d\mu(x_i).\\
& = \frac 1 {n!}\sum_{\substack{\tau, \tau' \in \mathfrak{S}_n \\ \tau(i) =\tau'(i) \forall i >2}} (-1)^\tau (-1)^{\tau'} \int_{E^2} 1_{x_1=x_2}\prod_{i=1}^2 \overline{\psi_{\tau(i)}}(x_i) \psi_{\tau'(i)}(x_i) d\mu(x_i).                \end{split}
                \end{equation*}
We see that such quantity is zero, because for every admissible $\tau$, $\tau'\in \mathfrak{S}_n$, the pair $\tau \circ \sigma_{12}$, $\tau'$ is also admissible, where $\sigma_{12}$ denotes the transposition of the elements $1,2$, and gives precisely the opposite contribution.

Let then $B_1$, \ldots,  $B_k \in \mathcal{E}$ be pairwise disjoint, and set $B_0 = E$. As we already argued that the process is simple, we can  rewrite the random variables $\sharp \left( \mathcal{X}_\rho \cap B_i\right)$ a.s.\ as follows:
\[
\sharp \left( \mathcal{X}_\rho \cap B_i\right) =  \sum_{\ell =1}^n  \chi_{B_i} \circ X_{\rho, \ell} \quad \text{for $i=1,\ldots, k$}
\]
and therefore, given $(m_i)_{i=1}^k$ with $\sum_i m_i = m \le n$, defining
\[
\mathcal{L} :=\left\{\ell:\left\{1, \ldots, n\right\} \to \left\{0,1, \ldots, k\right\}\, : \, \sharp \ell^{-1}(i) = m_i \quad \forall i =1,\ldots, k\right\},
\]
we find that
\[
\prod_{i=1}^k { \sharp \left(\mathcal{X}_\rho \cap B_i\right)  \choose m_i }= \sum_{\ell \in \mathcal{L}} \prod_{i=1}^n \chi_{B_{\ell(i)}} \circ X_{\rho, i}.
\]
Since the random variables $(X_{\rho,i})_{i=1,\ldots, n}$ have an exchangeable law, after taking expectation we can discuss only the case where $\prod_{i=1}^n B_{\ell(i)} = B_1^{m_1} \times B^{m_2} \times \ldots \times E^{n-m}$ and then multiply by their number
\[
\sharp \mathcal{L} = \frac{ n!}{m_1! \ldots, m_k! (n-m)!}.
\]
Using \eqref{eq:exp-slater-f}, we find therefore
\begin{equation*}\begin{split}
  \mathbb{E}&\left[\prod_{i=1}^m { \sharp \left( \mathcal{X}_\rho \cap B_i\right)  \choose m_i } \right] = \\
 & =\frac{  \sharp \mathcal{L} }{n!} \sum_{\tau, \tau' \in \mathfrak{S}_n} (-1)^\tau (-1)^{\tau'} \int_{ B_{1}^{m_1} \times \ldots B_k^{m_k}\times  E^{n-m} }   \prod_{i=1}^n \overline{\psi_{\tau(i)}}(x_i) \psi_{\tau'(i)}(x_i) d\mu(x_i)\\
 & = \frac{  \sharp \mathcal{L} }{n!}  \sum_{\tau(i) = \tau'(i)\,  \forall i=m+1, \ldots, n} (-1)^\tau (-1)^{\tau'} \int_{B_{1}^{m_1} \times \ldots B_k^{m_k}}  \prod_{i=1}^m \overline{\psi_{\tau(i)}}(x_i) \psi_{\tau'(i)}(x_i) d\mu(x_i),
 \end{split}
\end{equation*}
having integrated first with respect to the variables $x_{m+1}, \ldots, x_{n}$, and used the orthogonality of the states $\left\{\psi_j\right\}_{j=1}^n$, so  that only the permutations $\tau$, $\tau' \in \mathfrak{S}_n$ with $\tau(i) = \tau'(i)$ for $i=m+1, \ldots, n$ contribute. Defining $\sigma$ so that $\tau = \tau' \circ \sigma$, we have $\sigma(i) = i$ for $i=m+1, \ldots, n$, hence one can identify $\sigma$ with an element of $\mathfrak{S}_m$. Moreover, $(-1)^{\sigma} = (-1)^{\tau}(-1)^{\tau'}$, and
\[
\prod_{i=1}^m \psi_{ \tau'(i) }(x_i) = \prod_{i=1}^m \psi_{\tau'(\sigma(i))} (x_{\sigma(i)}) = \prod_{i=1}^m \psi_{\tau(i)}(x_{\sigma(i)}).
\]
Hence,
\begin{equation*}\begin{split}
 \sum_{\tau(i) = \tau'(i)\,  \forall i=m+1, \ldots, n} (-1)^\tau (-1)^{\tau'}  &  \prod_{i=1}^m \overline{\psi_{\tau(i)}}(x_i) \psi_{\tau'(i)}(x_i) =\\
 & =  \sum_{\tau \in \mathfrak{S}_n} \sum_{\sigma \in \mathfrak{S}_m} (-1)^\sigma  \prod_{i=1}^m \overline{\psi_{\tau(i)}}(x_i) \psi_{\tau(i)} (x_{\sigma(i)})\\
 & =  \sum_{\tau \in \mathfrak{S}_n}  \det \left(\left(\overline{\psi_{\tau(i)}}(x_i) \psi_{ \tau(i ) }(x_j)\right)_{i,j=1,\ldots, m}\right).
 \end{split}
\end{equation*}
To conclude, we argue that
\begin{equation}\label{eq:final-claim-determinant}
\frac{  \sharp \mathcal{L}}{n!}  \sum_{\tau \in \mathfrak{S}_n} \det\left(\left(\overline{\psi_{\tau(i)}}(x_i) \psi_{ \tau(i ) }(x_j)\right)_{i,j=1,\ldots, m}\right) = \det(K_\rho(x_i,x_j)_{i,j=1,\ldots, m}).
\end{equation}
Indeed, starting from the definition of $K_\rho$ and using the multilinearity of the determinant, we have
\begin{equation*}
\begin{split}  \det(K_\rho(x_i,x_j)_{i,j=1,\ldots, m})& = \det\left(\left(\sum_{\ell=1}^n \overline{\psi_\ell}(x_i) {\psi_\ell}(x_j)\right)_{i,j=1,\ldots, m}\right)\\
& = \sum_{\ell_1, \ldots, \ell_m=1}^n \det \left(\left(\overline{\psi_{\ell_i}}(x_i) \psi_{ \ell_i }(x_j)\right)_{i,j=1,\ldots, m}\right),\end{split}
\end{equation*}
Next, we can restrict the summation over $m$-uples $(\ell_1, \ldots, \ell_m)$ that have all different elements, otherwise the matrix $\left( \overline{\psi_{\ell_i}}(x_i) \psi_{ \ell_i }(x_j)\right)_{i,j=1,\ldots, m}$ has two (or more) rows that coincide, hence zero determinant. Finally, given any such $m$-uple $(\ell_1, \ldots, \ell_m)$, there exists $(n-m)!$ permutations $\tau \in \mathfrak{S}_n$ such that $\ell_i = \tau(i)$ for $i=1\ldots, m$, and \eqref{eq:final-claim-determinant} follows. \qedhere
\end{proof}

We are now in a position to prove \cref{prop:determinant-distance}.
\begin{proof}[Proof of \cref{prop:determinant-distance}]
We apply \cref{lem:slater-measure} to the Slater determinant state operators $\rho =\left|\Psi\right \rangle\left \langle\Psi\right|$, $\rho' =\left|\Psi'\right \rangle\left \langle\Psi'\right|$ associated to the families $\left\{\left|\psi_i\right \rangle\right\}_{i=1}^n$, $\left\{\left|\psi'_i\right \rangle\right\}_{i=1}^n$, so that the resulting random variables $\chi := \chi_\rho = \left\{X_{\rho, i}\right\}_{i=1}^n$, $\chi' := \chi_{\rho'} = \left\{X_{\rho', i}\right\}_{i=1}^n$ are determinantal point processes with kernels $K$, $K'$. Then, \eqref{eq:tv-determinantal} follows from \eqref{eq:contraction-trace-probabilities} and \eqref{eq:w1-determinantal} follows from \eqref{eq:contraction-W-random-variables}, that yield immediately
\[
\operatorname{TV}(X_\rho, X_{\rho'}) \le \sqrt{ 1- \left| \det\left(\left<\psi_i | \psi_j'\right>_{i,j=1\ldots, n}\right)\right|^2}
\]
and
\[
 W_{1}\left(X_{\rho}, X_{\rho'}\right) \le  n \sqrt{ 1 - \max_{V\in \mathfrak{U}_\Psi, U \in \mathfrak{U}_{\Psi'}}\left| \frac 1 n \sum_{i=1}^n\left<V \psi_{i} | U\psi_{i}'\right>\right|^2}.
\]
To conclude, we argue that
\[
\operatorname{TV}(\chi, \chi') \le   \operatorname{TV}(X_\rho, X_{\rho'}) \quad \text{and} \quad  W_{\#}(\chi, \chi') \le  W_1(X_\rho, X_{\rho'}).
\]
Indeed, the first inequality trivially follows from the fact that the total variation distance is a contraction with respect to any transformation, and in particular the map $(x_i)_{i=1}^n \in E^n \mapsto \left\{x_i\right\}_{i=1}^n$, which we use to define $\chi$ starting from $X_\rho$, and similarly $\chi'$ from $X_{\rho'}$. But the same map is a contraction also when we endow the space of subsets of $E$ with $n$ elements with the distance $\#$, and $E^n$ with the Hamming distance:
\[
\# \left(\left\{x_i\right\}_{i=1}^n, \left\{x'_i\right\}_{i=1}^n\right) = \sharp \left( \left\{x_i\right\}_{i=1}^n \Delta \left\{x'_i\right\}_{i=1}^n\right) \le \sum_{i=1}^n 1_{\left\{x_i \neq x'_i\right\}},
\]
showing also the validity of the second  inequality.
\end{proof}
Combining \cref{prop:determinant-distance} with the coupling argument from \cite[Theorem 13]{decreusefond2021optimal}, we obtain distance bounds for general determinantal point processes.

\begin{theorem}\label{thm:determinant-distance}
Let $\mathcal{X}$, $\mathcal{X}'$ be determinantal point processes on a common state space $(E, \mathcal{E})$ with respect to the same intensity measures $\mu$, and kernels
\begin{equation}\label{eq:Kern}
K = \sum_{i=1}^\infty \lambda_i \left|\psi_i\right\rangle\left \langle\psi_i\right|, \quad K'= \sum_{i=1}^\infty \lambda_i' \left|\psi_i'\right\rangle\left \langle\psi_i'\right|,
\end{equation}
with  $\lambda_i$, $\lambda_i' \in [0,1]$ such that $\sum_{i} \lambda_i < \infty$, $\sum_{i} \lambda_i'< \infty$ and orthonormal $\left\{\left|\psi_i\right\rangle \right\}_{i} \subseteq L^2(E, \mu)$, $\left\{\left|\psi_i'\right\rangle \right\}_{i} \subseteq L^2(E, \mu)$. Then, it holds
\begin{equation}\label{eq:determinantal-tv-general} \operatorname{TV}(\mathcal{X}, \mathcal{X}') \le \sum_{i=1}^\infty |\lambda_i-\lambda_i'| + \sum_{I}   \sqrt{ 1- \left| \det\left(\left<\psi_i | \psi_j'\right>_{i,j\in I}\right)\right|^2}w(\lambda, \lambda', I). \end{equation}
and
\begin{equation}\label{eq:determinantal-w-general}\begin{split} W_{\#}(\mathcal{X}, \mathcal{X}') & \le \left(2 + \sum_{i=1}^\infty \lambda_i + \sum_{i=1}^\infty \lambda_i'\right) \left(\sum_{i=1}^\infty |\lambda_i-\lambda_i'|\right)^{1/2} \\
& \quad  + \sum_{I}   \sharp I \sqrt{ 1 - \max_{V\in \mathfrak{U}_{ (\psi_i)_{i\in I} }, U \in \mathfrak{U}_{(\psi'_i)_{i \in I}} }\left| \frac 1 {\sharp I} \sum_{i\in I}\left<V \psi_{i} | U\psi_{i}'\right>\right|^2}w(\lambda, \lambda', I),\end{split}\end{equation}
where in both expressions, summation over $I$ is performed over all finite subsets of $\mathbb{N}\setminus \left\{0\right\}$ and we set
\begin{equation*}
 w(\lambda, \lambda', I) := \prod_{i \in I} \min\left\{\lambda_i, \lambda_i'\right\}\prod_{i \notin I}(1- \max\left\{\lambda_i, \lambda_i'\right\}).
\end{equation*}
\end{theorem}

\begin{remark}
 In the spectral decompositions of $K$ and $K'$ we do not require the eigenvalues to be ordered, hence one may further minimize the right hand sides in \eqref{eq:determinantal-tv-general} and \eqref{eq:determinantal-w-general} over all the bijections $\sigma$ of $\mathbb{N}\setminus \left\{0\right\}$. We also remark that the condition  $\sum_{i} \lambda_i < \infty$, i.e. $K$ is of trace class, is equivalent to a finite expected number of points for the point process, i.e., $\mathbb{E}[\sharp \mathcal{X}]<\infty$, and similarly for $\lambda_i'$ and $\mathcal{X}'$.
\end{remark}
\begin{proof}
Consider independent Bernoulli random variables $B = (B_i)_{i=1}^\infty$, with $P(B_i=1) = \lambda_i$, $P(B_i'=1) =\lambda_i'$ and define the random set $\mathcal{I} = \left\{i \, : \, B_i = 1\right\}$, and similarly $\mathcal{I}' =\left\{i\, : \, B_i' =1\right\}$. It is known that a process with the same law as $\mathcal{X}$ can be obtained by sampling, on each event $\mathcal{I} = I$, a determinantal point process $\mathcal{X}_I$, on $E$, with projection kernel $K_I = \sum_{i \in I} \left|\psi_i\right\rangle\left \langle\psi_i\right|$ and independent of $(B_i)_{i=1}^\infty$. A similar argument can be applied to $\mathcal{X}'$, i.e. by defining Bernoulli variables $B' = (B_i')_{i=1}^\infty$, the random set $\mathcal{I}'$ and determinant point processes $\mathcal{X}'_I$, with kernel $K_{I'} = \sum_{i \in I} \left|\psi_i'\right\rangle\left \langle\psi_i'\right|$, independent of $B'$, on the event $\mathcal{I}' =I$. Therefore, to define a coupling between $\mathcal{X}$ and $\mathcal{X}'$ it is sufficient to consider first a coupling between the projection a coupling between the two Bernoulli families $B$ and $B'$, which thus induces a coupling between the random sets $\mathcal{I}$, $\mathcal{I}'$ and, independently, a coupling between the determinantal point processes $\mathcal{X}_I$, $\mathcal{X}'_I$, for every $I \subseteq \left\{1, 2 \ldots \right\}$. Given such a coupling, recalling that the $\operatorname{TV}$ cost is the Wasserstein distance with respect to the trivial distance, we find
\begin{equation*}
\begin{split}
\operatorname{TV}(\mathcal{X}, \mathcal{X}') & \le P(\mathcal{X} \ne \mathcal{X}')  \le P(\mathcal{I} \neq \mathcal{I}') + \sum_{I}P(\mathcal{X}_I \neq \mathcal{X}_I' ) P(\mathcal{I} = \mathcal{I'} = I).
\end{split}
\end{equation*}
To find a simple expression in terms of the eigenvalues, we may simply build a coupling between $B$ and $B'$ by coupling each $B_i$ with $B_i'$ so that $P(B_i \neq B_i') = \operatorname{TV}(B_i, B_i') = |\lambda_i - \lambda_i'|$, so that
\[
P(B_i = B_i' = 1) = \min\left\{\lambda_i, \lambda_i'\right\}, \quad P(B_i = B_i' = 0) = 1 - \max\left\{\lambda_i, \lambda_i'\right\}
\]
and different $i$ give independent variables, thus
\[
P(B \neq B') \le \sum_{i} |\lambda_i - \lambda_i'|
\]
and for any given $I$,
\[
P( \mathcal{I} = \mathcal{I}' = I)  = \prod_{i \in I} \min\left\{\lambda_i, \lambda_i'\right\} \prod_{i \notin I}(1- \max\left\{\lambda_i, \lambda_i'\right\}) = w(\lambda, \lambda', I).
\]
Furthermore,  we can use for each $\mathcal{X}_I$, $\mathcal{X}_I'$ a coupling that minimizes the $\operatorname{TV}$ distance, hence by \eqref{eq:tv-determinantal} we bound from above
\begin{equation*}
 P(\mathcal{X}_I \neq \mathcal{X}_I' ) \le \sqrt{ 1- \left| \det\left(\left<\psi_i | \psi_j'\right>_{i,j\in I}\right)\right|^2}.
\end{equation*}
Combining these bounds, we find \eqref{eq:determinantal-tv-general}. Similarly, for the $W_\#$ cost, we find
\begin{equation*}
 \begin{split}
W_\#(\mathcal{X}, \mathcal{X}') & \le \mathbb{E}\left[\sharp (\mathcal{X} \Delta \mathcal{X}')\right] \\
& \le \mathbb{E}\left[\max\left\{\sharp \mathcal{I}, \sharp \mathcal{I}'\right\} I_{\mathcal{I} \neq \mathcal{I}'} \right] + \sum_{I}\mathbb{E}\left[\sharp ( \mathcal{X}_I \Delta  \mathcal{X}_I')  \right] P(\mathcal{I} = \mathcal{I'} = I).
 \end{split}
\end{equation*}
For the first term, we simply use Cauchy-Schwarz inequality
\begin{equation*}
\begin{split} \mathbb{E}\left[\max\left\{\sharp \mathcal{I}, \sharp \mathcal{I}'\right\} I_{\mathcal{I} \neq \mathcal{I}'} \right]& \le \mathbb{E}\left[\sharp \mathcal{I} I_{\mathcal{I} \neq \mathcal{I}'}\right] + \mathbb{E}\left[\sharp \mathcal{I} I_{\mathcal{I} \neq cI'} I_{\mathcal{I} \neq \mathcal{I}'} \right]\\
& \le\left(\mathbb{E}\left[(\sharp \mathcal{I})^2 \right]^{1/2} + \mathbb{E}\left[(\sharp \mathcal{I}')^2 \right]^{1/2} \right)P(B \neq B')^{1/2}\\
& \le \left(2 + \sum_{i=1}^\infty \lambda_i + \sum_{i=1}^\infty \lambda_i'\right) \left( \sum_{i=1}^\infty |\lambda_i-\lambda_i'|\right)^{1/2},
\end{split}
\end{equation*}
having used that
\[
\mathbb{E}\left[(\sharp I )^2\right]= \mathbb{E}\left[\left( \sum_{i} B_i\right)^2\right]\le \sum_{i}\lambda_i + \left(\sum_i \lambda_i\right)^2 \le \left(1+ \sum_{i}\lambda_i\right)^2.
\]
For the second term, we argue as in the total variation case, using instead a coupling that minimizes the $W_\#$ distance between $\mathcal{X}_I$ and $\mathcal{X}_I'$:
\begin{equation*}
 \mathbb{E}[\sharp ( \mathcal{X}_I \Delta  \mathcal{X}_I')  ] P(\mathcal{I} = \mathcal{I'} = I) \le  w(\lambda, \lambda', I)  \sharp I \sqrt{ 1 -\max_{V\in \mathfrak{U}_{ (\psi_i)_{i\in I} }, U \in \mathfrak{U}_{(\psi'_i)_{i \in I}} }\left| \frac 1 {\sharp I} \sum_{i\in I}\left<V \psi_{i} | U\psi_{i}'\right>\right|^2}\qedhere
\end{equation*}
\end{proof}

\section{Conclusion}\label{sec:conclusion}

By investigating the distances between quantum Slater determinant states and their classical counterparts, we demonstrate how quantum structures influence and constrain determinantal point processes. The rigorous bounds derived from trace and Wasserstein distances serve as a bridge linking quantum geometry to classical stochastic models. Our results indicate possible directions for future research:

\begin{enumerate}[label=\emph{\alph*})]
 \item Questions remain regarding the sharpness of the inequalities involved, in particular when equality in \eqref{eq:wass-slater} hold, and the existence of closed formulas for the quantum Wasserstein distance of order $1$ between Slater determinant states. Further, theoretical work could derive conditions under which sharpness in our equalities is achieved and analyze the implications for  quantum  systems.
 \item Future investigations may also explore algorithmic approaches to efficiently compute or approximate these distances, such as gradient descent methods or variational algorithms that leverage the structure of the unitary matrices.
 \item In the context of quantum many-body systems and quantum chemistry, Slater determinant states serve as a foundational tool for approximating the ground states of Hamiltonians with interacting terms. These states, while effective for non-interacting fermions, often fall short in accurately capturing the complexities introduced by interactions. Future research may investigate whether Wasserstein bounds can provide insights in improving or quantifying these approximations.
 \item Exploring alternative definitions of Wasserstein distances \cite{maas2024optimal, ding2025lower, beatty2025order, trevisan2025quantum} could pave the way to deriving bounds between determinantal point processes that take into account the geometric aspects of the configurations, possibly establishing a correct version of \eqref{eq:decreusefond-wrong}.
 \item  Our results apply specifically to fermionic quantum systems, where antisymmetry yields a determinantal structure. Extensions to other symmetry sectors, such as the bosonic case, seem to be non-trivial. Similarly, extending to spin-adapted configuration state functions would require additional tools and we leave it as as a natural direction for future work.
 \end{enumerate}

\section{Declarations.}

\paragraph{\textbf{Funding statement:}} D.T.\ and F.P.\ acknowledge the project  G24-202 ``Variational methods for geometric and optimal matching problems'' funded by Università Italo Francese.

D.T.\ acknowledges the MUR Excellence Department Project awarded to the Department of Mathematics, University of Pisa, CUP I57G22000700001,  the HPC Italian National Centre for HPC, Big Data and Quantum Computing -  CUP I53C22000690001, the PRIN 2022 Italian grant 2022WHZ5XH - ``understanding the LEarning process of QUantum Neural networks (LeQun)'', CUP J53D23003890006, the INdAM-GNAMPA project 2025 ``Analisi spettrale, armonica e stocastica in presenza di potenziali magnetici''.  Research also partly funded by PNRR - M4C2 - Investimento 1.3, Partenariato Esteso PE00000013 - "FAIR - Future Artificial Intelligence Research" - Spoke 1 "Human-centered AI", funded by the European Commission under the NextGeneration EU programme. Part of this research was performed while D.T.\ was visiting the Institute for Pure and Applied Mathematics (IPAM), which is supported by the National Science Foundation (Grant No. DMS-1925919).

C.B. acknowledges funding from the Italian Ministry of University and Research and Next Generation EU through the PRIN 2022 project PRIN202223CBOCC\_01, project code 2022AKRC5P. C.B. also acknowledges support of Grant PID2024-156184NB-I00 funded by MICIU/AEI/10.13039/501100011033 and cofunded by the European Union. C.B. warmly acknowledges also GNFM (Gruppo Nazionale per la Fisica Matematica) - INDAM.

F.P. acknowledges the grant "Progetti di ricerca d'Ateneo 2024" by Sapienza University, CUP B83C24006550001 and CUP B83C24007080005 and also acknowledges GNFM - INdAM.

\paragraph{\textbf{Competing interests:}} The authors have no competing interests to declare that are relevant to the content of this article.

\paragraph{\textbf{Data availability statement:}}Not applicable.

\paragraph{\textbf{Ethical standards:}} The research meets all ethical guidelines, including adherence to the legal requirements of the study country.

\paragraph{\textbf{Author contributions:}} All authors contributed equally.

\paragraph{\textbf{Supplementary material:}} No supplementary material is provided.

\end{document}